\documentstyle[12pt,supercite]{article}
\title{
\vspace{-2.5cm}
\hfill \parbox{2.3cm}{\small KOMA-96-30\\ July 1996}\\
\vspace*{1.6cm}
Computer Simulations of Quantum Chains}
\author{\large Wolfhard Janke$^{1}$ and Tilman Sauer$^{2}$ \\[3mm]
\em $^1$\,Institut f\"ur Physik,  \\
\em Johannes Gutenberg-Universit\"at Mainz, \\
\em 55099 Mainz, Germany  \\
\em $^2$\,Institut f\"ur Theoretische Physik, \\
\em Freie Universit\"at Berlin, \\
\em 14195 Berlin, Germany}
\date{}
\begin{document}
\maketitle
\begin{abstract}
We report recent progress in computer simulations of quantum systems
described in the path-integral formulation. For the example of the
$\phi^4$ quantum chain we show that the accuracy of the simulation
may greatly be enhanced by a combination of multigrid update techniques
with a refined discretization scheme.
This allows us to assess the accuracy of a variational approximation.
\end{abstract}
%
%-------------------------------------------------------------------
       \section{Introduction}
%-------------------------------------------------------------------
%
Monte Carlo (MC) simulations of many-particle quantum systems
based on a path-integral representation of the partition function
provide a numerical approach to these systems which in principle is free
of any approximations \cite{mc}.
The difficulty here is to achieve sufficient accuracy. Standard path-integral
MC simulations suffer from well-known draw backs, such as
systematic errors due to the necessary discretization and severe slowing down 
in the continuum limit. These problems clearly ask for the development of 
refined simulation techniques.

A Fourier MC simulation of the Sine-Gordon quantum chain was 
tried some time ago \cite{wr87}.
Preliminary data seemed to be quite promising,
but the method did not work quite so well for the related $\phi^4$ chain.
The reason is that the Fourier MC method is
not based on importance sampling which is a problem 
particularly for unbounded potentials such as the $\phi^4$ double well.

In view of these difficulties it is gratifying that recently 
algorithmic improvements developed for spin systems and 
lattice field theories could successfully be transferred to 
path-integral simulations \cite{js92b}.
Multigrid techniques and the staging algorithm 
have been shown to eliminate slowing down in the continuum limit for 
one-particle systems \cite{js93a,js96}. 
It seemed therefore worthwhile to investigate whether these 
refinements may now render simulations of quantum chains
sufficiently accurate to allow for a significant test of
analytical approximations. 
%
%------------------------------------------------------------------
       \section{Simulation techniques}
%------------------------------------------------------------------
%
The $N$-particle quantum systems we want to investigate are 
defined in any dimension D by the partition function
\begin{equation}
    {\cal Z} = e^{-\beta F} \!= \prod_{i=1}^N\int
               \!{\cal D}[\{\phi_i(u)\}]
               \exp\left[
                -\int_0^{\hbar\beta} \!\!du
                 \left(Aa\sum_{i=1}^N \frac{1}{2}\dot{\phi}_i^2(u)
                + V(\{\phi_i(u)\})\right)\!/\hbar\right],
 \label{eq:4partitionfunction}
\end{equation}
where $\beta=1/k_BT$ is the inverse temperature,
$\dot{\phi}_i\equiv d\phi_i/du$, and ${\cal D}[\{\phi_i(u)\}]$
denotes the functional measure for periodic paths, 
$\phi_i(0)=\phi_i(\hbar\beta)$.

For a simulation of these systems the partition function
(\ref{eq:4partitionfunction})
needs to be discretized. Standard discretization schemes
based on the Trotter formula 
$e^{A+B}=\lim_{L\rightarrow\infty}[e^{A/2L}e^{B/L}e^{A/2L}]^L$
for operators $A$ and $B$
entail a systematic error of the order $\epsilon^2$
where $\epsilon\equiv \hbar\beta/L$ and $L$ is the Trotter number.
A more rapidly converging discretization scheme of the order $\epsilon^4$
was proposed by Takahashi and Imada \cite{ti}. 
The only modification with respect to the standard discretization is that
the potential $V$ is replaced by an ``improved'' potential, 
\begin{equation}
   V_{\rm TI}(\{\phi_{i,k}\}) =   
             V(\{\phi_{i,k}\}) + \frac{\beta^2\hbar^2}{24AaL^2}
             \sum_{i=1}^N \left(\frac{\partial V}{\partial
                      \phi_{i,k}}\right)^2,
 \label{eq:4Vdiscrete}
\end{equation}
where $k$ denotes the additional index for the Trotter discretization
at each site. 

As far as the statistical error of the simulations is concerned
we expect a quadratic slowing down in the continuum limit
of large Trotter numbers $L$ 
for standard local update algorithms such as the Metropolis algorithm
\cite{js92b,js93a,js96}.
It is therefore desirable to apply refined update techniques 
which reduce autocorrelations in the MC process.
Fortunately, even though the discretized partition function effectively
represents a (D+1)-dimensional classical system, in many applications it is 
sufficient to apply one-dimensional
refined update schemes since we are approaching
the continuum limit only in the Trotter direction.
We may therefore use improved update schemes
developed for one-particle systems
at each site along the discretized time axis such as
one-dimensional multigrid cycles \cite{js93a} or the 
staging algorithm \cite{js96}.

The observables of interest are the internal energy and the specific
heat.
It is well-known that the definition of the internal energy $U$ 
gives rise to an estimator $U_k$ of the energy 
by differentiating the discretized partition function,
$U = - \partial \ln {\cal Z}/\partial \beta
                       \approx \overline{U_{\rm k}}$,
where $\overline{U_{\rm k}}$ denotes the arithmetic mean
over $N_m$ measurements of $U_{\rm k}$ in the MC process.
Applying a simple scaling argument one can also derive a different
but equivalent energy estimator $U_v$
based on the virial theorem with the same mean but different variance.
In order to further reduce the statistical error of the energy estimation
we may then use an optimized linear combination of these two 
estimators \cite{jstobepub}.
Note that the energy estimators differ for the standard
and the improved discretization schemes since the 
correction term in $V_{\rm TI}$ is $\beta$-dependent.
For the estimation of the specific heat analogous considerations apply.
A full account of the technical details discussing various
systematic algorithmic refinements of path-integral MC simulations 
will be given elsewhere \cite{jstobepub}.
%
%-------------------------------------------------------------------
       \section{Results for the $\phi^4$ quantum chain}
%-------------------------------------------------------------------
%
In the following we will present simulation results for the 
$\phi^4$ quantum chain \cite{js95a} where the potential $V$ in 
(\ref{eq:4partitionfunction}) 
is given by
\begin{equation}
      V(\{\phi_i\}) = Aa \sum_{i=1}^N \left[
                \frac{\omega_0^2}{2}(\phi_i-\phi_{i-1})^2
               + \frac{\omega_1^2}{8} (\phi_i^2-1)^2 \right].
 \label{eq:4potential}
\end{equation}
The partition function describes a set of $N$ harmonically coupled
oscillators of mass $Aa$  moving in double-well potentials separated 
by a distance $a$.
Following Ref.~\cite{gtv88b} we assume periodic boundary
conditions, $\phi_0\equiv\phi_N$, and introduce dimensionless parameters 
$R=\omega_0/\omega_1$, $Q = \hbar\omega_1/E_s$, and $t\equiv k_BT/E_s$, 
where $E_s=(2/3)Aa\omega_0\omega_1$ is the energy of 
the classical static kink and $R$ is its length in units of the lattice 
spacing $a$. The latter two parameters were kept fixed at $E_s=1$ and $R=5$.
The coupling constant $Q$ controls the 
quantum character of the system by determining whether the kinks are 
``heavy'' enough (small $Q$) to be treated semiclassically.

In our simulations, the number of oscillators was $N=300$ except for 
$t=0.05$, $0.30$, $0.35$, and $0.40$ where we simulated a chain of 
$N=200$ oscillators. The Trotter number was $L=16$ for $t\geq 0.20$, 
$L=32$ for $t=0.15$, $L=64$ for $t=0.10$, and $L=128$ for $t=0.05$. 
In all simulations we used the improved discretization scheme 
(\ref{eq:4Vdiscrete}).
The update was performed using a multigrid W-cycle with piecewise
constant interpolation in Trotter direction at each site
with single-hit Metropolis updating and $n_1=1$ pre-, $n_2=0$ 
postsweeps \cite{js93a,js95a}.
The thermodynamic observables of interest are the
internal energy per site, $u = U/N$
and the specific heat per site given by
$c = C/N = \partial u/\partial T$.
More precisely, we will be interested only in the anharmonic
contribution to these quantities. For the free energy this is given
by $dF\equiv F-F_{\rm harmon} = F-(1/\beta)\sum_{k=1}^N\ln(2\sinh F_k)$,
where $F_k=\beta\hbar\omega_k/2$ and
$\omega_k^2=4\omega_0^2\sin^2(k\pi/N)+\omega_1^2$.
For each data point we have measured the internal energy
using the optimally combined estimator with
$N_m=200\,000$ measurements taken every second sweep, 
after discarding $2\,000$ sweeps for thermalization.
The Metropolis acceptance rates were adjusted to be $\approx 40-60\%$
on the finest grid and the same step width was used for all
multigrid levels.
The specific heat was measured by simple numerical differentiation
of the ``combined'' estimator which was reweighted 
in a temperature interval of $dt=0.0001$.
These estimates gave consistent values with direct measurements
of the specific heat using the estimators obtained by differentiating
the discrete partition function but (slightly) smaller errors.
All statistical errors were computed by jack-kniving the data on the basis
of $500$ blocks.

In Fig.~1 we compare our simulation data with analytical results
based on a variational ansatz \cite{gt}
which has been shown to be a very powerful and useful method
for an approximate evaluation of quantum partition 
functions \cite{gtv88b,gt,janke}.
The variational approach starts from a quadratic trial Hamiltonian
whose parameters are determined
by optimizing the Jensen-Peierls inequality for the free energy.
A numerical solution of the resulting set of $N(N+1)/2$
self-consistent equations is extremely complicated and
only the limiting cases
of high and low temperatures and for small coupling $Q$ have been
treated in the literature.  This adds another
source of error to the uncertainty inherent in the
variational ansatz itself.
For the $\phi^4$ quantum chain \cite{gtv88b}
our simulations thus provide independent data for an assessment of
the accuracy of this approach.

\begin{figure}[t]
\vskip  6.0 truecm
\includegraphics{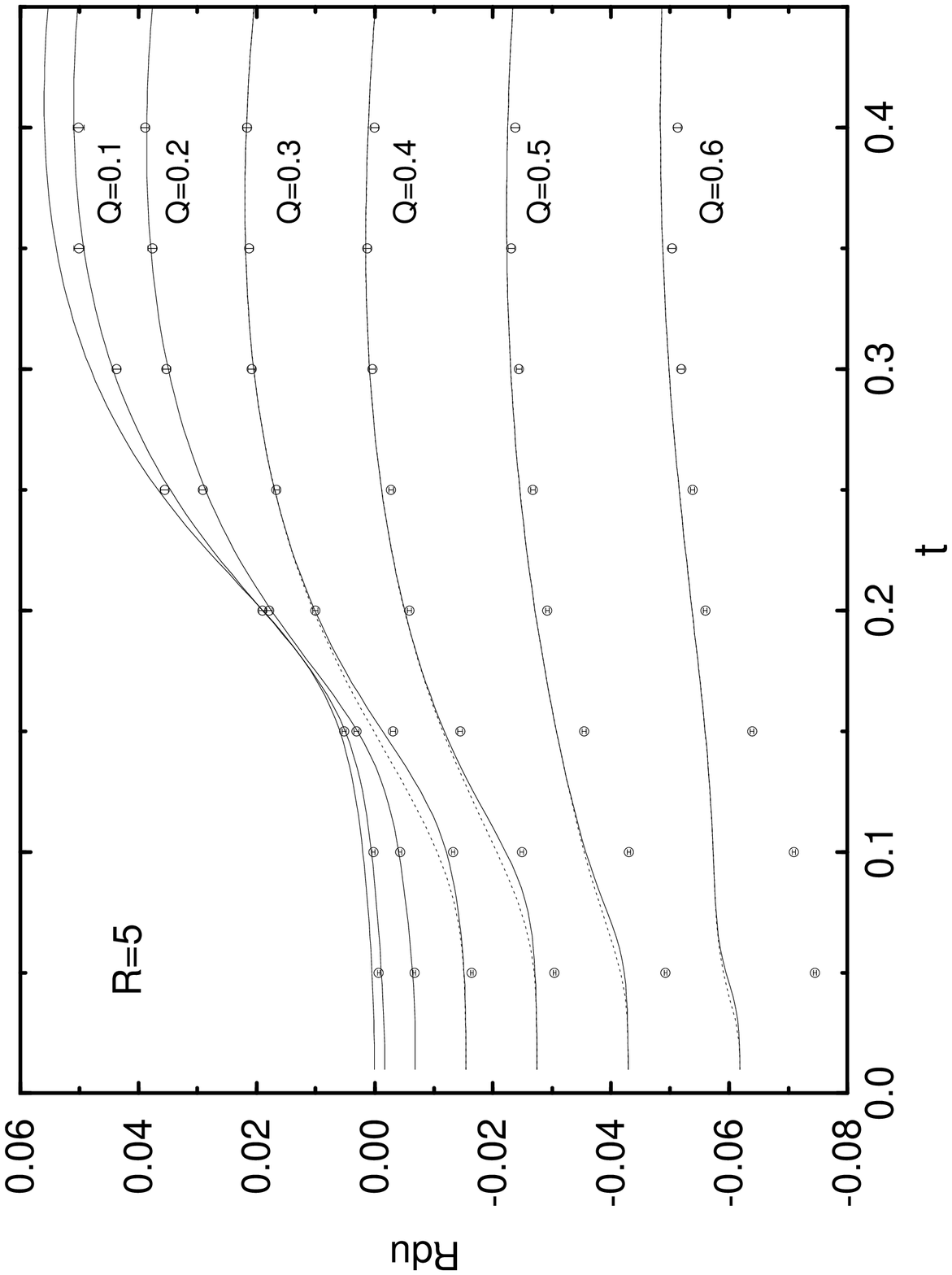}
\includegraphics{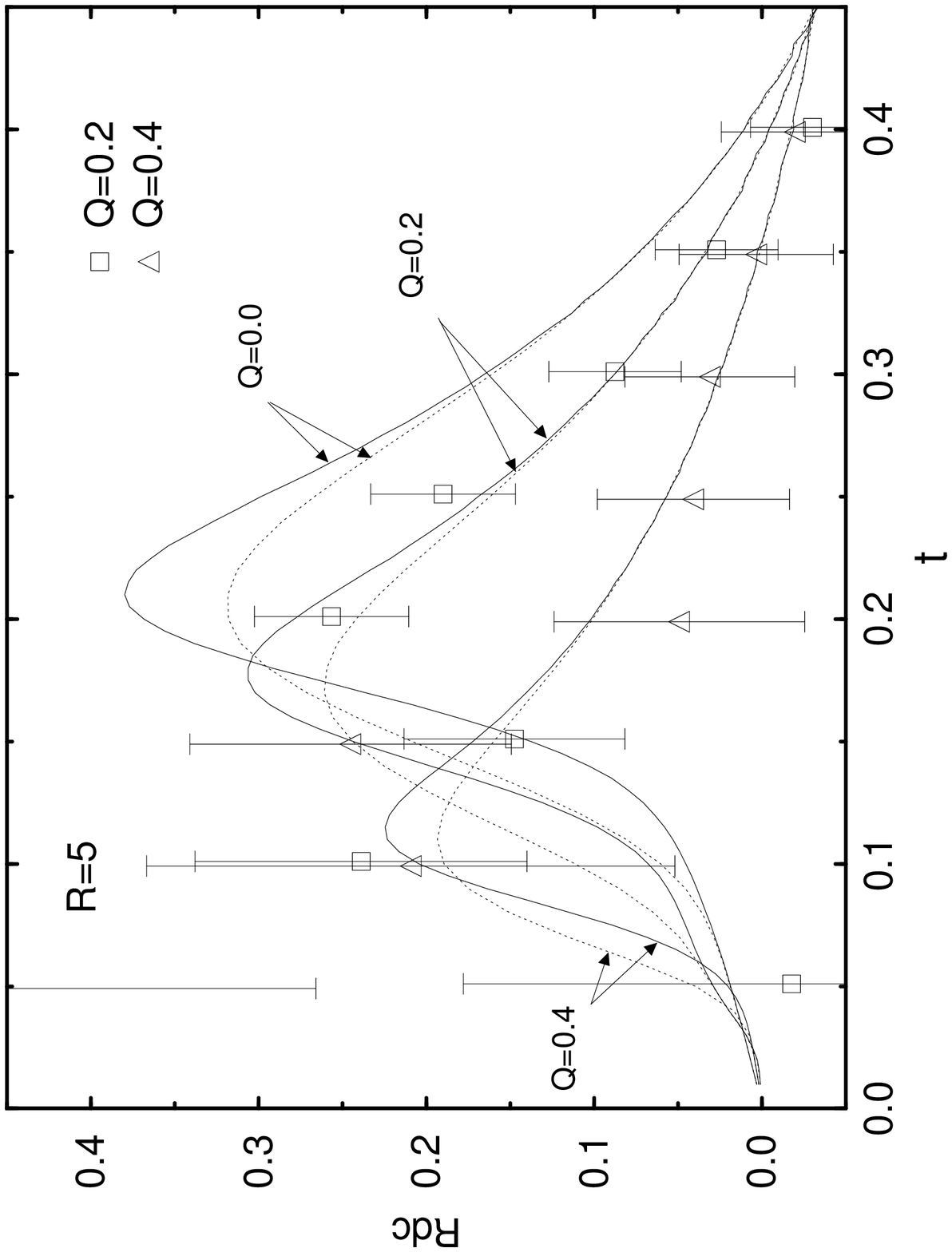}
\caption[Anharmonic contributions to the internal energy $Rdu$ as
a function of temperature $t$ for various couplings $Q$]{%
Anharmonic contributions to the internal energy and specific heat per site.
Solid (dotted) lines show the variational approximation for $N=300$
($N=\infty$).
}
\label{figure:Rdu}
\end{figure}
Let us first look at the temperature dependence of the internal energy.
Figure~\ref{figure:Rdu} shows the measured anharmonic contributions
to the internal energy per site as a function of the temperature $t$ for
various couplings $Q$.
Regarding a comparison with the variational data we observe that
the approximation is fully confirmed for high temperatures $t$
and small couplings $Q$.
For lower $t$ we still find a satisfactory agreement 
if we also take into account finite-size corrections in the
variational approach by evaluating the effective classical
partition function given in eq.~(4.6) of
Ref.~\cite{gtv88b} for finite $N$ with the help of the 
transfer matrix method \cite{js95a}.

The situation is different, however, for low temperatures 
and large couplings as can be clearly seen in Fig.~\ref{figure:Rdu}.
Here we find significant deviations from the variational approximation.
Note that the error bars for the data are
smaller than the data symbols.
Let us take a closer look at the lowest temperature which we
have investigated, $t=0.05$. 
For $Q=0.1$ and $Q=0.2$ the variational approximation is still
confirmed within the statistical errors. But already for $Q=0.3$ we find 
a statistically significant discrepancy which further increases 
when we go to larger couplings. For the worst
case, $Q=0.6$, the variational approximation 
deviates from the MC results already by $56$ 
error bars, even if finite-size corrections are fully taken
into account.
We emphasize that within the statistical errors our data are to be
regarded exact, i.e. that the deviations are entirely due to the
variational approximation.

Let us finally take a look at the plot on the r.h.s.\ of 
Fig.~\ref{figure:Rdu} which shows the measured anharmonic 
contributions to the specific heat per site.
Due to the fact that the estimation of the specific heat involves
a difference of statistically fluctuating variables the resulting
statistical accuracy is greatly reduced compared to the
estimation of energies. Therefore the accuracy of
our data for the specific heat unfortunately still does not allow for
a significant falsifying test of the variational approximation.
%
%------------------------------------------------------------------
       \section{Conclusions}
%------------------------------------------------------------------
%
By employing refined path-integral MC techniques we have been able to 
drastically reduce the systematic and statistical errors of a quantum
MC simulation of the $\phi^4$ chain. The resulting accuracy now allows
for a significant test of the variational approximation. For small 
couplings $Q$ we find very good agreement. Only for large couplings and low
temperatures do we observe significant deviations from the MC data. The 
discrepancies may be due to the limitations inherent in the variational
approximation or to the additional approximation of the small coupling 
expansion. It would therefore be interesting to see whether the MC data
can be reproduced by evaluating the variational equations to higher
orders of $Q$ or by
taking into account higher-order corrections to the
variational approach itself \cite{kleinert93a}.\\[-0.15cm]

W.J. thanks the Deutsche Forschungsgemeinschaft for a Heisenberg
fellowship.
\vspace*{-0.20cm}
%
%-----------------------------------------------------------------------
               
%                          
\end{document}